\documentclass{SCIS2026}
\usepackage{algorithm}
\usepackage{algpseudocode}
\usepackage{caption}

\begin{document}
\ArticleType{LETTER}
\Year{2025}
\Month{January}
\Vol{68}
\No{1}
\DOI{}
\ArtNo{}
\ReceiveDate{}
\ReviseDate{}
\AcceptDate{}
\OnlineDate{}
\AuthorMark{}
\AuthorCitation{}

\title{Offline dedicated quantum attacks on block cipher constructions based on two parallel permutation-based pseudorandom functions}{Offline dedicated quantum attacks on block cipher constructions based on two parallel permutation-based pseudorandom functions}

\author[1,2]{Xiao-Fan Zhen}{}
\author[2]{Zhen-Qiang Li}{lizq@sklc.org}
\author[1]{Jia-Cheng Fan}{}
\author[1]{Su-Juan Qin}{}
\author[1,2]{Fei Gao}{gaof@bupt.edu.cn}


\address[1]{State Key Laboratory of Networking and Switching Technology, Beijing University of Posts and Telecommunications,\\ Beijing, 100876, China}
\address[2]{State Key Laboratory of Cryptology, P. O. Box 5159, Beijing 100878, China}

\maketitle

\begin{multicols}{2}
\noindent Quantum computing poses unprecedented threats to cryptographic security. Simon's algorithm~\cite{simon1997power} and Grover-meets-Simon algorithm~\cite{leander2017grover} have become important tools in the quantum cryptanalysis of symmetric cryptography. The resulting attacks generally rely on superposition queries to the encryption oracle (i.e., the Q2 model), which limits their practical relevance. To remove this limitation, Bonnetain {\it et al.}~\cite{bonnetain2019quantum} introduced the \emph{offline} Simon's algorithm, which decouples the queries to the encryption oracle from the Grover iterations, allowing the attack to be performed using either superposition queries or classical queries (i.e., the Q1 model). Recently, Shi {\it et al.}~\cite{shi2024dedicated} proposed a dedicated quantum attack on XOR-type functions in the Q2 model, applied it to several BBB MACs, and showed that it simultaneously reduces the circuit depth, width, and gate count compared with the parallel Grover-meets-Simon algorithm.
However, two questions remain open: whether their attack applies to other cryptographic constructions, and whether it can be performed in the offline setting (particularly using only classical queries to the encryption oracle).

To address these two questions, we make the following contributions.
\begin{enumerate}
    \item We identify new cryptographic structures amenable to Shi {\it et al.}'s attack, namely PolyMAC and block cipher constructions based on Two Parallel Permutation-based Pseudorandom Functions (TPP-PRFs), including $\mathsf{XopEM}$, $\mathsf{SoEM22}$, $\mathsf{SUMPIP}$, and $\mathsf{DS\text{-}SoEM}$, thereby answering Shi {\it et al.}'s open question~\cite{shi2024dedicated}.
    \item We construct decoupled XOR-type functions and propose an \emph{offline} quantum attack on constructions based on TPP-PRFs, which breaks the obstacle that Shi {\it et al.}'s attack relies on \emph{superposition} queries. Compared to previous results, our attack reduces the query complexity from $O(2^{(n+t)/2}\cdot (n-t))$ to $O(2^{t}\cdot (n-t))$ in the Q2 model, where $t$ is a truncation parameter satisfying $0<t<n$ and $n$ is the input length. Further, we reduce both the time complexity and classical query complexity from $\tilde O(2^{2n/3})$ to $\tilde O(2^{(2n-t)/3})$ in the Q1 model for $0<t \leq n/2$ (for example, see $\mathsf{SoEM22}$ in Table~\ref{tab:soem22}).
\end{enumerate}

\lettersection{New instances amenable to Shi {\it et al.}'s attack}
Shi {\it et al.}~\cite{shi2024dedicated} proposed a dedicated quantum attack on XOR-type functions of the form $f(i, x) = g_1(x) \oplus g_2(x \oplus \alpha_i)$, where $f(i,\cdot)$ is periodic for the correct value of $i$. We show that PolyMAC is vulnerable to this attack by constructing the XOR-type function
\begin{equation}
f(i,x)=\text{PolyMAC}(\beta_{0}, x)\oplus \text{PolyMAC}\left(\beta_{1}, x\oplus \alpha_i\right),
\end{equation}
where $\beta_0,\beta_1\in\{0,1\}^n$ are fixed strings. The function $f$ has a period $s=(k_1\oplus  k_3)\odot (\beta_0 \oplus \beta_1)$ when $\alpha_i=k_1\odot (\beta_{0}\oplus \beta_{1})$ or $k_3\odot (\beta_{0}\oplus \beta_{1})$. This identifies PolyMAC as a new application of Shi {\it et al.}'s attack.

Moreover, constructions based on TPP-PRFs also admit such XOR-type functions and are therefore amenable to this attack. A brief description of the attack is given in Appendix~A.

\lettersection{Obstacle to offline conversion and a way around it}
For the above attack on PolyMAC, the calculation of both
$g_1(x)=\text{PolyMAC}(\beta_{0}, x)$ and
$g_2(x\oplus \alpha_i)=\text{PolyMAC}(\beta_{1}, x\oplus \alpha_i)$ require quantum access to PolyMAC. Thus, for XOR-type functions in which both components depend on the encryption oracle, including the PolyMAC instance above and those derived from BBB MACs~\cite{shi2024dedicated}, each Grover iteration of Shi {\it et al.}'s attack still requires quantum access to the encryption oracle.

This suggests looking for XOR-type functions whose queries to the encryption oracle can be decoupled from the Grover iterations. We next consider functions of the form
\begin{equation}
f(i,x)=g_1(x)\oplus p(x\oplus \alpha_i),
\end{equation}
where $p$ is a public function that can be computed offline and only $g_1$ requires access to the encryption oracle. We call this form a \emph{decoupled XOR-type function}, or simply a $p$-XOR-type function. To simplify the presentation throughout the remainder of this paper, we assume $\alpha_i=i$, so that
$f(i,x)=g_1(x)\oplus p(x\oplus i)$.

\lettersection{Construction of $p$-XOR-type functions from TPP-PRFs}
We show that TPP-PRFs can be used to construct $p$-XOR-type functions. A TPP-PRFs takes an $n$-bit input $x$ and two parallel permutations $P_{1}$ and $P_{2}$, and is defined by
$g(x)=l_{33}P_1\big(l_{13}(x) \oplus l_{14}(k_1)\big)\oplus l_{34}P_2\big(l_{23}(x) \oplus l_{24}(k_2)\big)\oplus e(x)\oplus C$,
where $e(x)=l_{31}l_{11}(x) \oplus l_{32}l_{21}(x)$ and $C$ is a key-dependent constant.
\end{multicols}
\begin{center}
\footnotesize
\captionof{table}{Previous and new quantum attacks on $\mathsf{SoEM22}$. $\#$Query: queries to $\mathsf{SoEM22}$. $\#$Time: offline computation time. $0<t< n$ in the Q2 model, and $0<t \leq n/2$ in the Q1 model. The best results are in bold.}
\label{tab:soem22}
\setlength{\tabcolsep}{4pt}
\renewcommand{\arraystretch}{1.25}
\begin{tabular*}{\textwidth}{@{\extracolsep{\fill}}ccccccc}
\toprule
Algorithm & Model & Iteration & \#Query & \#Time & Qubit & Reference\\ \hline
Grover-meets-Simon & Q2 & $2^{n/2}$ & $O(2^{n/2}\cdot n)$ & $O(2^{n/2}\cdot n^{3} )$ & $O(n^2)$ &~\cite{leander2017grover}\\
Dedicated attack & Q2 & $2^{(n-t)/2}$ & $O\big(2^{(n+t)/2} \cdot (n-t)\big)$ & $O(2^{(n-t)/2} \cdot n^3)$ & $\tilde O(2^t)$ &~\cite{shi2024dedicated},  Ours \\
Offline dedicated attack & Q2 & $2^{(n-t)/2}$ & $\mathbf{O\big(2^{t}\cdot (n-t)\big)}$ & $O(2^{(n-t)/2} \cdot n^3)$ & $\tilde O(2^t)$ & Ours \\
Offline Simon & Q1 & $2^{2n/3}$ & $O(2^{2n/3})$ & $O(2^{2n/3}\cdot n^{3})$ & $O(n^2)$  & ~\cite{bonnetain2019quantum,sun2025quantum}\\
Offline dedicated attack & Q1 & $2^{(2n-t)/3}$ & $\mathbf{O(2^{(2n-t)/3})}$ & $\mathbf{O(2^{(2n-t)/3}\cdot n^3)}$ & $\tilde O(2^t)$  & Ours \\
\bottomrule
\end{tabular*}
\end{center}
\begin{multicols}{2}
Based on this TPP-PRF, we construct the function
\begin{equation}\label{eq:TPP-PRFs-f}
    f(i,x)= g(x) \oplus e(x)\oplus l_{33}P_{1}\left(l_{13}(x) \right)\oplus l_{34}P_{2}\left(l_{23}(x\oplus i)\right),
\end{equation}
which means $g_1(x)=g(x)\oplus e(x) \oplus l_{33}P_1(l_{13}(x))$ and $p(x\oplus i)=l_{34}P_2(l_{23}(x\oplus i))$.
The function $f(i,x)$ satisfies $f(i,x)=f(i,x\oplus l_{13}^{-1}l_{14}(k_1))$ when $i=l_{23}^{-1}l_{24}(k_2)$ or $i=l_{23}^{-1}l_{24}(k_2)\oplus l_{13}^{-1}l_{14}(k_1)$. Since $i$ is XORed with $x$ in the public function $P_{2}$, $f$ is a $p$-XOR-type function.

To derive the form needed for the offline attack, we further apply Shi {\it et al.}'s truncation technique to Eq.~\eqref{eq:TPP-PRFs-f} and obtain
\begin{equation}\label{eq:TPP-PRFs-F}
F^{\mathcal L}(i^l,x^l)=\bigoplus\limits_{u\in \mathcal L}f(i^l\| 0^{t}, x^{l} \| 0^{t} \oplus u)=G_1^{\mathcal L}(x^l)\oplus P^{\mathcal L}(i^l,x^l),
\end{equation}
where $G_1^{\mathcal L}(x^l)=\bigoplus\limits_{u\in\mathcal L}g_1(x^l\|0^t\oplus u)$ and $P^{\mathcal L}(i^l,x^l)=\bigoplus\limits_{u\in\mathcal L}p\big((x^l\oplus i^l)\|0^t\oplus u\big)$. Then $F^{\mathcal{L}}(i^l,x^l)$ has an $(n-t)$-bit period $s^l=l_{13}^{-1}l_{14}k_1^{l}$ for $i^l=i^l_{0}$, where $i^l_{0}=l_{23}^{-1}l_{24}k_2^l$ or $i^l=l_{23}^{-1}l_{24}k_2^l\oplus l_{13}^{-1}l_{14}k_1^l$.

Several block cipher constructions, including $\mathsf{XopEM}$, $\mathsf{SoEM22}$, $\mathsf{SUMPIP}$, and $\mathsf{DS\text{-}SoEM}$, are instantiations of TPP-PRFs. Thus, the above analysis yields the corresponding $p$-XOR-type functions and truncated conditional periodic functions for these constructions. The details are given in Appendix~B.

\lettersection{Offline dedicated quantum attack on block ciphers based on TPP-PRFs}
Based on Eqs.~\eqref{eq:TPP-PRFs-f} and~\eqref{eq:TPP-PRFs-F}, we adapt the offline Simon framework to Shi {\it et al.}'s attack to propose offline dedicated quantum attacks on TPP-PRF-based constructions in the Q1 and Q2 models.
The key observation is that the decoupled property of $p$-XOR-type functions allows the superposition state $|\psi_{G_1^{\mathcal L}}\rangle= \bigotimes^{c'} \left( \sum_{x^l \in \{0,1\}^{n-t}} \vert x^{l} \rangle |G_1^{\mathcal{L}}(x^l)\rangle \right)$ to be prepared first, and the remaining computations become independent of the encryption oracle.
Accordingly, our attack has two phases.
\begin{enumerate}
    \item Online phase: prepare the state $|\psi_{G_1^{\mathcal L}}\rangle$ requiring $O(2^n)$ classical queries in the Q1 model or $O\big(2^t\cdot (n-t)\big)$ quantum queries in the Q2 model.
    \item Offline phase: starting from the prepared state $|\psi_{G_1^{\mathcal L}}\rangle$, the test procedure is applied to $F^{\mathcal L}(i^l,x^l)$ to check whether it has a hidden period.  After the test, the state $|\psi_{G_1^{\mathcal L}}\rangle$ is restored and can be reused in subsequent iterations.
\end{enumerate}
The complete descriptions of two phases in the Q1 and Q2 models are provided in Appendix~C.
The complexity of the offline dedicated attack is summarized in Theorem~\ref{th:p-xor}.
\begin{theorem}\label{th:p-xor}
Let \( f: \{0, 1\}^{\kappa} \times \{0, 1\}^{n} \to \{0, 1\}^{m} \) be a decoupled XOR-type function, there exists an offline dedicated quantum attack on block cipher constructions based on TPP-PRFs, which recovers $i_0^l\in \{0,1\}^{\kappa-t}$ with $O(2^{(\kappa-t)/2} \cdot n^3)$ time and $\tilde O(2^t)$ qubits, requiring $O(2^n)$ classical queries to the encryption oracle in the Q1 model, or $O(2^t\cdot (n-t))$ quantum queries in the Q2 model, where $0<t<n$, \( c' = n + \kappa - 2t + \tau + 1 \).
\end{theorem}

Taking $\mathsf{SoEM22}$ as an example, we construct the $p$-XOR-type function $f(i,x)=\mathsf{SoEM22}(x)\oplus P_{1}(x)\oplus P_{2}(x \oplus i)$, which satisfies $f(i,x)=f(i,x\oplus k_1)$ when $i=k_2$ or $i=k_2\oplus k_1$.

In the Q2 model, we obtain
$F^{\mathcal{L}}(i^{l}, x^{l})=\bigoplus\limits_{u\in \mathcal{L}}\mathsf{SoEM22}(x^l\|0^t\oplus u)
\oplus P_{1}(x^l\|0^t\oplus u) \oplus P_{2}\left((x^l \oplus i^{l})\|0^t\oplus u\right)$, which has an $(n-t)$-bit period $k^{l}_{1}$ when $i^l=k^{l}_{2}$ or $i^l=k^{l}_{2}\oplus k^{l}_{1}$.
According to Theorem~\ref{th:p-xor}, we recover either \( k^{l}_{2}\) or $k^{l}_{2}\oplus k^{l}_{1}$ with $O\big(2^t\cdot (n-t)\big)$ quantum queries to the $\mathsf{SoEM22}$ oracle and \( O(2^{(n-t)/2}\cdot n^3) \) offline computation time.

In the Q1 model, we construct the function $F^{\mathcal{L}}(i^l\|j^{l_2},x^{l_1})=\bigoplus\limits_{u\in \mathcal{L}}\mathsf{SoEM22}(x^{l_1}\|0^{n-t-p}\|0^{t}\oplus u)\oplus P_{1}(x^{l_1}\|j^{l_2}\|0^{t}\oplus u)\oplus P_{2}\Big(\big((x^{l_1}\|0^{n-t-p})\oplus i^{l}\big)\|0^t\oplus u\Big)$, which has a $p$-bit period $k_{1}^{l_1}$ when $i^l\|j^{l_2}=k_{2}^l\|k_{1}^{l_2}$ or $i^l\|j^{l_2}=k_{2}^l \oplus (k_{1}^{l_1}\|0^{n-t-p}) \|k_{1}^{l_2}$.
We can recover one of $k_2^l\|k_1^{l_2}$ and $k_{2}^l \oplus (k_{1}^{l_1}\|0^{n-t-p}) \|k_{1}^{l_2}$ by applying Theorem~\ref{th:p-xor}, requiring $O(2^{p+t})$ classical queries to the $\mathsf{SoEM22}$ oracle and $O(2^{(2n-2t-p)/2}\cdot n^3)$ offline computation time. Let $D$ be the number of classical queries and $T$ the offline computation time. This attack achieves a tradeoff of $T^2D=2^{2n-t}$, which is balanced at $T=D=\tilde O(2^{(2n-t)/3})$ for $0<t \leq n/2$. The remaining key bits can be recovered using the offline Simon's algorithm with negligible additional cost.

A comparison with previous attacks on $\mathsf{SoEM22}$ is given in Table~\ref{tab:soem22}. The offline dedicated attacks on general TPP-PRFs are presented in Appendix~D.

\lettersection{Conclusion}
In this work, we identified PolyMAC and block cipher constructions based on TPP-PRFs as new instances vulnerable to dedicated quantum attacks.
By defining a decoupled XOR-type function, we further proposed offline dedicated quantum attacks on constructions based on TPP-PRFs, with lower query complexity than previous attacks in both the Q1 and Q2 models.
Our results provide new insights into the quantum security of block ciphers based on pseudorandom functions and motivate further investigation into more generic and effective offline dedicated quantum attacks applicable to other cryptographic structures.
\Acknowledgements{This work was supported by the National Natural Science Foundation of China (Grant Nos. U25B2014, 62372048, 62272056, 62371069).}

\Supplements{Appendix A-D.}

\end{multicols}
\ArticleType{Supplementary File}

\clearpage
\begin{appendix}

\section{Dedicated attacks on PolyMAC and constructions based on TPP-PRFs}
PolyMAC scheme~\cite{kim2020tight} is a Double-block Hash-then-Sum construction based on polynomial evaluation, proposed in 2020. It uses two hashing keys $ k_{1},k_{3}\in\{0,1\}^{n} $ and two encryption keys $ k_{2},k_{4}\in\{0,1\}^{m} $. We consider the case with the two-block message.
Throughout this section, $\odot$ denotes multiplication in $GF(2^n)$

\begin{equation}
\text{PolyMAC}(m_{1},m_{2}) = E_{k_{2}}(k_{1}^{2}\odot m_{1} \oplus k_{1}\odot m_{2}) \oplus E_{k_{4}}(k_{3}^{2}\odot m_{1} \oplus k_{3}\odot m_{2}),
\end{equation}
where $M=m_1\|m_2$, and $|m_1|=|m_2|=n$.

Based on PolyMAC, we construct the XOR-type function. Let $\beta_0$, $\beta_1$ be two fixed strings in $\{0,1\}^{n}$, $\beta_0\neq \beta_1$, and $\alpha_i\in \{0,1\}^{n}$, we define the function as
\begin{equation}
\begin{aligned}\label{eq:polymac}
f: \{0,1\}^{n} \times \{0,1\}^{n} & \rightarrow \{0,1\}^{n} \\
(\alpha_i, x) & \mapsto \text{PolyMAC}(\beta_{0}, x)
\oplus \text{PolyMAC}\left(\beta_{1}, x\oplus \alpha_i\right).
\end{aligned}
\end{equation}
Since the secret state $\alpha_i$ is XORed with $x$ in the second term $\text{PolyMAC}(\beta_1, x \oplus \alpha_i)$ of Eq.~\eqref{eq:polymac}, and the function has the period $s=(k_1\oplus  k_3)\odot (\beta_0 \oplus \beta_1)$ when $\alpha_i=k_1\odot (\beta_{0}\oplus \beta_{1})$ or $\alpha_i=k_3\odot (\beta_{0}\oplus \beta_{1})$. Then, the XOR-type function given in Eq.~\eqref{eq:polymac} satisfies the requirement of Shi {\it et al.}'s attack.

Now, we define a new function $F^{\mathcal{L}}$ using truncated input
\begin{equation}
\begin{aligned}\label{eq:polymacF}
F^{\mathcal{L}} : \{0,1\}^{n-t} \times \{0,1\}^{n-t}  &\rightarrow \{0,1\}^m  \\
(\alpha_{i}^{l}, x^{l}) & \mapsto \bigoplus_{u \in \mathcal{L}} f(\alpha_i^l\| 0^t, x^l \| 0^t \oplus u),
\end{aligned}
\end{equation}
where $\mathcal{L} = \{ u : u = 0^{n-t} \| * \}$, $\alpha_i = \alpha_i^l \| \alpha_i^r$, $x = x^l \| x^r$, $|\alpha_i^l| = |x^l| = n-t$, and $|\alpha_i^r| = |x^r| = t$.
We deduce that the function $F^{\mathcal{L}}$ has a period $ s^l=(k_1\oplus  k_3)^l\odot (\beta_0 \oplus \beta_1)^l$ when $\alpha_i^l=k_1^l\odot (\beta_{0}\oplus \beta_{1})^l$ or $k_3^l\odot (\beta_{0}\oplus \beta_{1})^l$. Applying Shi {\it et al.}'s attack~\cite{shi2024dedicated} to PolyMAC, we can recover the high $(n-t)$-bit $k_1^l\odot (\beta_{0}\oplus \beta_{1})^l$ or $k_3^l\odot (\beta_{0}\oplus \beta_{1})^l$ with $O\big(2^{(n+t)/2} \cdot (n-t)\big)$ quantum queries to the PolyMAC oracle and $O(2^{(n-t)/2} \cdot n^3)$ time.

The remaining $t$ bits can be recovered as follows. Taking $\alpha_i^l=k_1^l\odot (\beta_{0}\oplus \beta_{1})^l$ as an example, we define $f'(\alpha_i^r,x)=f\big(k_1^l\odot (\beta_{0}\oplus \beta_{1})^l\|\alpha_ i^r,x\big)$. When $\alpha_i^r=k_1^r\odot (\beta_{0}\oplus \beta_{1})^r$, $f'(\alpha_i^r,x)=f'(\alpha_i^r,x\oplus s)$, which requires $O(2^{t/2})$ quantum queries.

A TPP-PRF~\cite {chen2019build} is defined by
\begin{equation}\label{eq:PRFs-g}
g(x)=l_{33}P_1\big(l_{13}(x) \oplus l_{14}(k_1)\big)\oplus l_{34}P_2\big(l_{23}(x) \oplus l_{24}(k_2)\big)\oplus e(x)\oplus C,
\end{equation}
where $e(x)=l_{31}l_{11}(x) \oplus l_{32}l_{21}(x)$ and $C$ is a key-dependent constant.

For this TPP-PRF, we construct the function
\begin{equation}\label{eq:TPP-PRFs-f}
    f(i,x)= g(x) \oplus e(x)\oplus l_{33}P_{1}\left(l_{13}(x) \right)\oplus l_{34}P_{2}\left(l_{23}(x\oplus i)\right),
\end{equation}
which satisfies $f(i,x)=f(i,x\oplus l_{13}^{-1}l_{14}(k_1))$ when $i=l_{23}^{-1}l_{24}(k_2)$ or $i=l_{23}^{-1}l_{24}(k_2)\oplus l_{13}^{-1}l_{14}(k_1)$. Therefore, constructions based on TPP-PRFs also admit such XOR-type functions and are therefore amenable to Shi {\it et al.}'s attack.

Applying the truncation techniques, we obtain
\begin{equation}
\begin{aligned}\label{eq:TPP-PRFsF}
F^{\mathcal{L}} : \{0,1\}^{n-t} \times \{0,1\}^{n-t}  &\rightarrow \{0,1\}^m  \\
(\alpha_{i}^{l}, x^{l}) & \mapsto \bigoplus_{u\in \mathcal L}f(i^l\| 0^{t}, x^{l} \| 0^{t} \oplus u).
\end{aligned}
\end{equation}
Then $F^{\mathcal{L}}(i^l,x^l)$ has an $(n-t)$-bit period $s^l=l_{13}^{-1}l_{14}k_1^{l}$ for $i^l=i^l_{0}$, where $i^l_{0}=l_{23}^{-1}l_{24}k_2^l$ or $i^l=l_{23}^{-1}l_{24}k_2^l\oplus l_{13}^{-1}l_{14}k_1^l$. Similarly, Shi {\it et al.}'s attack is applied to constructions based on TPP-PRFs to recover $l_{23}^{-1}l_{24}k_2^l$ with $O\big(2^{(n+t)/2} \cdot (n-t)\big)$ quantum queries and $O(2^{(n-t)/2} \cdot n^3)$ time. The remaining key bits can then be recovered by Grover-meets-Simon algorithm.

\section{Construction of $p$-XOR-type functions from TPP-PRFs}
\subsection{General construction and proof of the period property}
For the $p$-XOR-type function constructed by TPP-PRFs,
the truncation in the Q2 model yields the same conditional periodic function as in Shi {\it et al.}'s attack, namely Eq.~\eqref{eq:TPP-PRFsF}. Then, we only need to define the corresponding function $F^{\mathcal L}$ in the Q1 model.
\begin{equation}
\begin{aligned}\label{eq:PRF-F}
F^{\mathcal L}(i^l\|j^{l_2}, x^{l_1})=&G_1^{\mathcal L}(x^{l_1})\oplus P^{\mathcal L}(i^l\|j^{l_2}, x^{l_1}),
\end{aligned}
\end{equation}
where $l_1\in \{0,1\}^{p}$, $l_2\in \{0,1\}^{n-t-p}$, $l=l_1\|l_2$.
The function $G_1^{\mathcal L}:\{0,1\}^p\rightarrow\{0,1\}^n$ and the public function $P^{\mathcal L}:\{0,1\}^{2n-2t-p}\times \{0,1\}^{p} \rightarrow  \{0,1\}^n$ are defined as
\begin{align*}
    G_1^{\mathcal L}(x^{l_1})= &\bigoplus \limits_{u\in \mathcal{L}}g(x^{l_1}\|0^{n-t-p}\|0^t\oplus u)\oplus e(x^{l_1}\|0^{n-t-p}\|0^t\oplus u),\\
    P^{\mathcal L}(i^l\|j^{l_2}, x^{l_1})= &\bigoplus \limits_{u\in \mathcal{L}}l_{33}P_{1}\left(l_{13}(x^{l_1}\|j^{l_2}) \|0^t\oplus u\right)\oplus l_{34}P_{2}\Big(l_{23}\big((x^{l_1}\|0^{n-t-p})\oplus i^l\big)\|0^t\oplus u\Big).
\end{align*}

The function $F^{\mathcal L}$ given in Eq.~\eqref{eq:PRF-F} has the period $s^{l_1}=l_{13}^{-1}l_{14}k_1^{l_1}$ when $i^l\|j^{l_2}=l_{23}^{-1}l_{24}k_2^l\|l_{13}^{-1}l_{14}k_1^{l_2}$ or $i^l\|j^{l_2}=l_{23}^{-1}l_{24}k_2^l\oplus (l_{13}^{-1}l_{14}k_1^{l_1}\|0^{n-t-p})\|l_{13}^{-1}l_{14}k_1^{l_2}$. Take $i^l\|j^{l_2}=l_{23}^{-1}l_{24}k_2^l\|l_{13}^{-1}l_{14}k_1^{l_2}$ as an example, we deduce that
\begin{align*}
&F^{\mathcal L}(l_{23}^{-1}l_{24}k_2^l\|l_{13}^{-1}l_{14}k_1^{l_2}, x^{l_1}) \\
=&\bigoplus \limits_{u\in \mathcal{L}}l_{33}P_1\big((l_{13}x^{l_1}\oplus l_{14}k_1^{l_1})\|l_{14}k_1^{l_2}\|0^t\oplus u\big) \oplus l_{34}P_2\big((l_{23}x^{l_1}\oplus l_{24}k_2^{l_1})\| l_{24}k_2^{l_2}\|0^t\oplus u\big)\\
\quad &\oplus l_{33}P_1\big(l_{13}(x^{l_1}\|l_{13}^{-1}l_{14}k_1^{l_2})\|0^t\oplus u\big) \oplus l_{34}P_2\big(l_{23}(x^{l_1}\|0^{n-t-p}\oplus l_{23}^{-1}l_{24}k_2^l)\|0^t\oplus u\big)\oplus C\\
\quad =&\bigoplus \limits_{u\in \mathcal{L}}l_{33}P_1(l_{13}x^{l_1}\|l_{14}k_1^{l_2}\|0^t\oplus u) \oplus l_{34}P_2\big((l_{23}x^{l_1}\oplus l_{23}l_{13}^{-1}l_{14}k_1^{l_1}\oplus l_{24}k_2^{l_1})\| l_{24}k_2^{l_2}\|0^t\oplus u\big)\\
\quad &\oplus l_{33}P_1\Big(\big(l_{13}(x^{l_1}\oplus l_{13}^{-1}l_{14}k_1^{l_1})\|l_{13}^{-1}l_{14}k_1^{l_2}\big)\|0^t\oplus u\Big) \oplus l_{34}P_2\Big(\big(l_{23}(x^{l_1}\oplus l_{13}^{-1}l_{14}k_1^{l_1})\|0^{n-t-p}\oplus l_{23}^{-1}l_{24}k_2^l\big)\|0^t\oplus u\Big)\oplus C\\
=&F^{\mathcal L}(l_{23}^{-1}l_{24}k_2^l\|l_{13}^{-1}l_{14}k_1^{l_2}, x^{l_1}\oplus l_{13}^{-1}l_{14}k_1^{l_1}).
\end{align*}

\subsection{Instantiations of TPP-PRFs}

The Xop construction~\cite{bellare1998luby} is defined by the bitwise XOR of the outputs from two distinct pseudorandom permutations (PRPs) applied to the same input $x$:
\begin{equation*}
    \text{Xop}_{E_{1},E_{2}}(x) = E_{1}(x) \oplus E_{2}(x),
\end{equation*}
where $E_{1}$ and $E_{2}$ represent the encryption algorithms of the respective PRPs.

Realizing the Xop construction with two Even-Mansour ciphers, $EM_{1}(x) = P_{1}(x \oplus k_{1}) \oplus k_{2}$ and
$EM_{2}(x) = P_{2}(x \oplus k_{3}) \oplus k_{4}$, resulting in the $\mathsf{XopEM}$ function:
\begin{equation}
\mathsf{XopEM}(x) = P_{1}(x \oplus k_{1}) \oplus P_{2}(x \oplus k_{3}) \oplus k_{2} \oplus k_{4}.
\end{equation}
We can obtain a $p$-XOR-type function as follows:
$f(i,x) =P_{1}(x \oplus k_{1}) \oplus P_{2}(x \oplus k_{3}) \oplus P_{1}(x) \oplus P_{2}(x \oplus i)$, where $g_1(x) = \mathsf{XopEM}(x)$, and the public function is $p(i,x)= P_1(x)\oplus P_{2}(x \oplus i)$. This function has a period $s=k_1$ precisely when $i=k_{3}$ or $i=k_{3}\oplus k_1$.

The function $F^{\mathcal{L}}: \{0,1\}^{n-t} \times \{0,1\}^{n-t-p} \times \{0,1\}^{p}\rightarrow \{0,1\}^{n}$ is defined as:
\begin{equation}
    \begin{aligned}\label{eq:XopEM-F}
    F^{\mathcal{L}}(i^l\|j^{l_2},x^{l_1})= \bigoplus\limits_{u\in \mathcal{L}}\mathsf{XopEM}(x^{l_1}\|0^{n-t-p}\|0^{t}\oplus u)\oplus P_{1}(x^{l_1}\|j^{l_2}\|0^{t}\oplus u)\oplus P_{2}\Big(\big((x^{l_1}\|0^{n-t-p})\oplus i^{l}\big)\|0^t\oplus u\Big).
\end{aligned}
\end{equation}
It can be verified that $F^{\mathcal{L}}(i^l\|j^{l_2},x^{l_1})=F^{\mathcal{L}}(i^l\|j^{l_2},x^{l_1}\oplus k_1^{l_1})$ when $i^l\|j^{l_2}=k_{3}^l\|k_{1}^{l_2}$ or $i^l\|j^{l_2}=k_{3}^l\oplus (k_{1}^{l_1}\|0^{n-t-p})\|k_{1}^{l_2}$.

\textbf{SoEM22~\cite{chen2019build}.} The Sum of Even-Mansour (SoEM) construction employs two public permutations $P_1$, $P_2$, and two $n$-bit keys $k_1$, $k_2$. It is a specific instance of the XOR construction where $k_3=k_2$ and $k_4=k_1$, defined as:
\begin{equation}
\mathsf{SoEM22}(x) = P_{1}(x \oplus k_{1}) \oplus P_{2}(x \oplus k_{2}) \oplus k_{1} \oplus k_{2}.
\end{equation}

We can construct a $p$-XOR-type function $f(i,x)=\mathsf{SoEM22}(x)\oplus P_1(x) \oplus P_2(x \oplus i)$, where $P_1(x) \oplus P_2(x \oplus i)$ is a public function. This function satisfies
$f(i,x)=f(i,x\oplus k_1)$ when $i=k_2$ or $i=k_2\oplus k_1$. Next, we construct the function $F^{\mathcal{L}}(i^l\|j^{l_2},x^{l_1})$ as:
\begin{equation}
    \begin{aligned}\label{Eq:Soem22Q1}
    F^{\mathcal{L}}(i^l\|j^{l_2},x^{l_1})=&\bigoplus\limits_{u\in \mathcal{L}}\mathsf{SoEM22}(x^{l_1}\|0^{n-t-p}\|0^{t}\oplus u)\oplus P_{1}(x^{l_1}\|j^{l_2}\|0^{t}\oplus u)\oplus P_{2}\Big(\big((x^{l_1}\|0^{n-t-p})\oplus i^{l}\big)\|0^t\oplus u\Big).
    \end{aligned}
\end{equation}
The function $F^{\mathcal{L}}(i^l\|j^{l_2},x^{l_1})$ given in Eq.~\eqref{Eq:Soem22Q1} has a period $k_{1}^{l_1}$ when $i^l\|j^{l_2}=k_{2}^l\|k_{1}^{l_2}$ or $i^l\|j^{l_2}=k_{2}^l \oplus (k_{1}^{l_1}\|0^{n-t-p}) \|k_{1}^{l_2}$.

If $P_1=P$ and $P_2=P^{-1}$, the above $\mathsf{SoEM22}$ construction corresponds to $\mathsf{SUMPIP}$~\cite{guo2019beyond}. In the same way, a $p$-XOR-type function $f(i,x)$ and a hidden periodic function $F^{\mathcal{L}}$ can be constructed based on the truncation parameter.

\textbf{DS-SoEM~\cite{bhattacharjee2022cencpp}.}
This construction is a domain-separated variant of the Xop construction that uses a $d$-bit domain-separation constant and is a sum of two Even-Mansour ciphers. For an input $x \in \{0,1\}^{n-d}$, it is defined as:
\begin{equation}
\begin{aligned}\label{eq:ds-soem}
\mathsf{DS\text{-}SoEM}(x) =&P\left( \left(x \oplus \mathrm{msb}_{n-d}(k_1)\right) \| 0^{d} \right) \oplus P\left( \left(x \oplus \mathrm{msb}_{n-d}(k_2)\right) \| 1^{d} \right) \oplus k_1 \oplus k_2,
\end{aligned}
\end{equation}
where `$\mathrm{msb}_{n-d}$' denotes truncation to the $n-d$ most significant bits.

We define a $p$-XOR-type function
$f(i,x) = \mathsf{DS\text{-}SoEM}(x)\oplus P(x\| 0^{d})\oplus P\left((x\oplus i)\| 1^{d}\right)$, where the public function is $P(x\| 0^{d})\oplus P\left((x\oplus i)\| 1^{d}\right)$. This function satisfies $f(i,x)=f\left(i,x\oplus \mathrm{msb}_{n-d}(k_1)\right)$ when $i=\mathrm{msb}_{n-d}(k_2)$ or $i=\mathrm{msb}_{n-d}(k_2)\oplus \mathrm{msb}_{n-d}(k_1)$. The new function $F^{\mathcal{L}}$ is defined as:
\begin{equation}
    \begin{aligned}\label{eq:ds-soem-F}
        F^{\mathcal{L}}(i^l\|j^{l_2},x^{l_1})=&
\bigoplus_{u \in \mathcal{L}} \mathsf{DS\text{-}SoEM}(x^{l_1} \| 0^{n-t-p}\| 0^t \oplus u)\oplus P\big((x^{l_1}\|j^{l_2}\|0^t\oplus u)\|0^d\big)\oplus P\Big(\big((x^{l_1}\|0^{n-t-p} \oplus i^{l})\|0^t\oplus u\big) \|1^d\Big).
    \end{aligned}
\end{equation}
This function has a $p$-bit period $[\mathrm{msb}_{n-d}(k_1)]^{l_1}$. It occurs when $i^l\|j^{l_2}=A$ or $i^l\|j^{l_2}=B$, where $A=[\mathrm{msb}_{n-d}(k_2)]^l\|[\mathrm{msb}_{n-d}(k_1)]^{l_2}$ and $B=[\mathrm{msb}_{n-d}(k_2)]^l\oplus \left([\mathrm{msb}_{n-d}(k_1)]^{l_1}\|0^{n-t-p}\right)\|[\mathrm{msb}_{n-d}(k_1)]^{l_2}$.

In summary, our analysis shows that several instances of TPP-PRF, including $\mathsf{XopEM}$, $\mathsf{SoEM22}$, $\mathsf{SUMPIP}$, and $\mathsf{DS\text{-}SoEM}$, can be used to construct $p$-XOR-type functions. Furthermore, by applying truncation techniques, we present their corresponding conditional periodic functions $F^{\mathcal{L}}$, as shown in Eqs.~\eqref{eq:XopEM-F},~\eqref{Eq:Soem22Q1}, and~\eqref{eq:ds-soem-F}.
These explicit forms of $F^{\mathcal L}$ will be used by the offline attack in Appendix~C and by the applications to general TPP-PRF-based constructions in Appendix~D.

\section{Algorithmic details of the offline dedicated quantum attack}
Based on the above $p$-XOR-type functions constructed by TPP-PRFs, we propose an offline dedicated quantum attack on TPP-PRF-based constructions in the Q1 and Q2 models (see Algorithm~\ref{al:p-xor}).
\begin{algorithm}
\footnotesize
\begin{algorithmic}[1]
\caption{Offline dedicated attack on block cipher constructions based on $p$-XOR-type function using truncated technique}
\label{al:p-xor}
\Require{
Prepared $\left|\psi_{G_1^{\mathcal{L}}}\right\rangle = \otimes^{c'} \left( \sum_{x^l \in \{0,1\}^{n-t}} \vert x^{l} \rangle |G_1^{\mathcal{L}}(x^l)\rangle \right) $,
\Statex Let $\mathcal{L}=\{u:u=0^{n-t}||*\}$, $f(i,x)$ be a $p$-XOR-type function, and $f(i,\cdot )$ have a non-trivial period $s$ when $i=i_{0}$. }
\Ensure The high $(\kappa-t)$-bit of $i_0$ and $(n-t)$-bit of $s$, namely $i_0^l$ and $s^l$, s.t., $F^{\mathcal{L}}(i^l,x^l)=F^{\mathcal{L}}(i^l,x^l\oplus s^l)$ when $i^l=i^l_0$.
\State Start $\left|\psi_{G_1^{\mathcal{L}}}\right\rangle |0^{\kappa-t}\rangle$.
\State Apply $I_{c'(n-t+m)}\otimes H^{\otimes \kappa-t}$ to obtain $|\Psi\rangle$:
\Statex \hfill $|\Psi\rangle = |\psi_{G_1^{\mathcal{L}}}\rangle \left(\sum_{i^l \in \{0,1\}^{\kappa-t}} |i^l\rangle\right)$. \hfill \null
\State Repeat $\mathcal O (2^{(\kappa-t)/2})$ Grover iterations:
\Statex \hfill $|i^l\rangle|b\rangle \overset{\text{test}}{\rightarrow} |v\rangle |b\oplus r\rangle$.\hfill
\null
\Statex Note that the test oracle is a unitary operator that takes $|\psi_{G_1^{\mathcal{L}}}\rangle |i^l\rangle$ as input, and tests whether $F^{\mathcal{L}}(i^l, \cdot)$ has a hidden period (see Algorithm~\ref{al:p-xor-test} for details).
\State Measure the index $i^l$ to obtain $i_{0}^{l}$.
\end{algorithmic}
\end{algorithm}

\begin{algorithm}
\footnotesize
\begin{algorithmic}[1]
\caption{The procedure checks whether a function $F^{\mathcal{L}}(i^l,x^l)=G_1^{\mathcal{L}}(x^l)\oplus P^{\mathcal{L}}(i^l,x^l)$ has a period, without making any new queries to $g_1$}
\label{al:p-xor-test}
\State Start $|\psi_{G_1^{\mathcal{L}}}\rangle|b\rangle$:
\Statex
\hfill $\left( \sum _{x_{1}^{l},\cdots ,x_{c'}^{l} \in \{0,1\}^{n-t}} \vert x_{1}^{l} \rangle\cdots \vert x_{c'}^{l} \rangle |G_1^{\mathcal{L}}(x_{1}^{l})\rangle\cdots |G_1^{\mathcal{L}}(x_{c'}^{l})\rangle \right)|b\rangle$. \hfill \null
\State Apply $c'$ $U_{P^{\mathcal{L}}}$ to obtain $|\psi_{F^{\mathcal{L}}}\rangle$:
\Statex \hfill$|\psi_{F^{\mathcal{L}}}\rangle=\otimes^{c'}  \left(\sum_{x^l \in \{0,1\}^{n-t}} |x^{l}\rangle|G_1^{\mathcal{L}}(x^{l}) \oplus P^{\mathcal{L}}(x^{l})\rangle \right)|b\rangle=\otimes^{c'}  \left(\sum_{x^l \in \{0,1\}^{n-t}} |x^{l}\rangle|F^{\mathcal{L}}(x^{l})\rangle \right)|b\rangle$.\hfill \null
\State Apply $(H^{\otimes (n-t)} \otimes I_m)^{\otimes c'} \otimes I_1$ to $|\psi_{F^{\mathcal{L}}}\rangle$, to get:
\Statex \hfill
$\left(\sum_{v_{1},x_{1}^{l}\in \{0,1\}^{n-t}}(-1)^{v_1\cdot x_{1}^{l}}|v_{1}\rangle |F^{\mathcal{L}}(x_{1}^{l})\rangle\right)\otimes\cdots\otimes\left(\sum_{v_{c'},x_{c'}^{l}\in \{0,1\}^{n-t}}(-1)^{v_{c'}\cdot x_{c'}^{l}}|v_{c'}\rangle |F^{\mathcal{L}}(x_{c'}^{l})\rangle\right)|b\rangle$.\hfill \null
\State Compute $d := \dim(\text{Span}(v_1, \dots, v_{c'}))$.
\If {$d = n-t$}
    \State set $r := 0$
\Else
    \State set $r := 1$, and add $r$ to $b$
\EndIf
\State Uncompute steps 3-2 to obtain $|\psi_{G_1^{\mathcal{L}}}\rangle|b \oplus r\rangle$. \hfill \null
\end{algorithmic}
\end{algorithm}
This algorithm is developed by integrating the offline Simon's algorithm with the dedicated quantum attack on block ciphers via the construction of $p$-XOR-type functions. Our algorithm is separated into two phases: the online phase prepares the superposition state $|\psi_{G_1^{\mathcal L}}\rangle$ via classical or quantum queries to the encryption oracle $U_{g_1}$ (requiring $O(2^n)$ classical queries in the Q1 model by Algorithm~\ref{al:p-xor-G}, or $O(2^t\cdot (n-t))$ quantum queries in the Q2 model by Algorithm~\ref{al:p-xor-G2}), and the offline phase performs the remaining quantum computations independently of the encryption oracle (using the test procedure in Algorithm~\ref {al:p-xor-test} to check whether the function $F^{\mathcal L}$ has a hidden period without any new queries to the encryption oracle). This separation achieves a reduction in query complexity, as Grover iterations are executed offline after a limited number of online interactions.
\begin{algorithm}
\footnotesize
\begin{algorithmic}[1] 
\caption{Prepare $|\psi_{G_1^{\mathcal{L}}}\rangle$ in the Q1 model} 
\label{al:p-xor-G}
\Require Classical query to the encryption oracle $g_1$.
\Ensure
$|\psi_{G_1^{\mathcal{L}}}\rangle =\otimes^{c'} \left( \sum_{x^l \in \{0,1\}^{n-t}} \vert x^{l} \rangle |G_1^{\mathcal{L}}(x^l)\rangle \right)$
\State Apply $(H^{\otimes (n-t)}\otimes I_{m})^{\otimes c'}$ to obtain:
\Statex \hfill$\sum\limits_{x_{1}^{l},\cdots ,x_{c'}^{l} \in \{0,1\}^{n-t}} |x_{1}^{l}\rangle \cdots |x_{c'}^{l} \rangle \vert 0^{c'm} \rangle$. \hfill
\null
\State For each $x^l\in \{0,1\}^{n-t}$, classical queries to $g_1$ to get $\bigoplus\limits_{u\in \mathcal{L}}g_1(x^{l}||0^{t}\oplus u)$.
\State Apply unitary operation, which writes $\bigoplus\limits_{u\in \mathcal{L}}g_1(x^{l}||0^{t}\oplus u)$ in the second register if the first contains the value $x^{l}$: \Statex
\hfill
$|\psi_{G_1^{\mathcal{L}}}\rangle =\otimes^{c'} \left( \sum_{x^l \in \{0,1\}^{n-t}} \vert x^{l} \rangle \vert \bigoplus\limits_{u\in \mathcal{L}}g_1(x^{l}||0^{t}\oplus u)\rangle \right) =\otimes^{c'} \left( \sum_{x^l \in \{0,1\}^{n-t}} \vert x^{l} \rangle |G_1^{\mathcal{L}}(x^l)\rangle \right)$.\hfill\null
\end{algorithmic}
\end{algorithm}

\begin{algorithm}
\footnotesize
\begin{algorithmic}[1] 
\caption{Prepare $\vert\psi_{G_1^{\mathcal{L}}}\rangle$ in the Q2 model} 
\label{al:p-xor-G2}
\Require quantum queries to the encryption oracle $g_1$.
\Ensure
$|\psi_{G_1^{\mathcal{L}}}\rangle =\otimes^{c'} \left( \sum_{x^l \in \{0,1\}^{n-t}} \vert x^{l} \rangle |G_1^{\mathcal{L}}(x^l)\rangle \right)$
\State Apply $(H^{\otimes (n-t)}\otimes I_{m})^{\otimes c'}$ to obtain:
\Statex \hfill$\sum\limits_{x_{1}^{l},\cdots ,x_{c'}^{l} \in \{0,1\}^{n-t}} |x_{1}^{l}\rangle \cdots |x_{c'}^{l} \rangle \vert 0^{c'm} \rangle$.\hfill
\null
\State  The XOR sum of $2^t$ distinct $U_{g_1}$, which can be implemented by using CNOTs, we can obtain $|\psi_{G_1^{\mathcal{L}}}\rangle $: \Statex
\hfill
$|\psi_{G_1^{\mathcal{L}}}\rangle=\otimes^{c'} \left( \sum_{x^l \in \{0,1\}^{n-t}} \vert x^{l} \rangle |G_1^{\mathcal{L}}(x^l)\rangle \right)$.\hfill\null
\end{algorithmic}
\end{algorithm}

We now explain the implementation of Algorithm~\ref{al:p-xor}. The goal is to recover the high $(\kappa-t)$ bits $i_0^l$ such that $F^{\mathcal{L}}(i_0^{l}, \cdot)$ has a non-trivial period $s^l$. Starting with the superposition state $|\psi_{G_1^{\mathcal{L}}}\rangle=\otimes^{c'} \left( \sum_{x^l \in \{0,1\}^{n-t}} \vert x^{l} \rangle |G_1^{\mathcal{L}}(x^l)\rangle \right)$, the input is $|\psi_{G_1^{\mathcal{L}}}\rangle |0^{\kappa-t}\rangle$ of $(\kappa-t) + c^{\prime}(n-t) + c^{\prime}m$ qubits. The initial state $\vert \Psi \rangle$ required for the Grover iteration is obtained by several $H$ transforms. The iterative function $G = D_{\vert \Psi \rangle} O_{\text{test}}$ is run for $\mathcal{O}(2^{(\kappa-t)/2})$ iterations, where $D_{| \Psi \rangle} = (2 | \Psi \rangle \langle \Psi \vert - I)$. The final step is measuring the index $i^{l}$ to obtain $i_0^{l}$. The implementation of $U_{P^{\mathcal{L}}}$ and  $O_{\text{test}}$ in Algorithm~\ref {al:p-xor-test} is similar to that in Ref.~\cite{shi2024dedicated}. Specifically, $U_{P^{\mathcal{L}}}$ implements the function $$P^{\mathcal{L}}(i^l, x^{l}) = \bigoplus_{u \in \mathcal{L}}  p\big((x^{l} \oplus i^{l}) \Vert 0^t \oplus u\big),$$ which is the XOR sum of $2^t$ public functions $p$ and can be implemented in parallel without a sequential relationship.

\section{Offline dedicated attacks on constructions based on TPP-PRFs}
Recall that the function $f(i,x)$ defined in Eq.~\eqref{eq:TPP-PRFs-f} is a $p$-XOR type function, with $g_1(x)=g(x) \oplus e(x)\oplus l_{33}P_{1}\big(l_{13}(x) \big)$, and the public function $p(i, x) = l_{34}P_{2}\big(l_{23}(x\oplus i)\big)$. Hence, our offline attack can be applied to constructions based on TPP-PRFs.

In the Q1 model, we construct a conditional periodic function $F^{\mathcal L}(i^l\|j^{l_2}, x^{l_1})$ given in Eq.~\eqref{eq:PRF-F} by the truncation techniques, which has a $p$-bit  period $s^{l_1}=l_{13}^{-1}l_{14}k_1^{l_1}$ when $i^l\|j^{l_2}=l_{23}^{-1}l_{24}k_2^l\|l_{13}^{-1}l_{14}k_1^{l_2}$ or $i^l\|j^{l_2}=l_{23}^{-1}l_{24}k_2^l\oplus (l_{13}^{-1}l_{14}k_1^{l_1}\|0^{n-t-p})\|l_{13}^{-1}l_{14}k_1^{l_2}$.
The offline attack recovers either $l_{23}^{-1}l_{24}k_2^l\|l_{13}^{-1}l_{14}k_1^{l_2}$ or $i^l\|j^{l_2}=l_{23}^{-1}l_{24}k_2^l\oplus (l_{13}^{-1}l_{14}k_1^{l_1}\|0^{n-t-p})\|l_{13}^{-1}l_{14}k_1^{l_2}$, requiring $O(2^{p+t})$ classical queries to the encryption oracle and $O(2^{(2n-2t-p)/2}\cdot n^3)$ offline computation time, where $0\leq p\leq n-t$, $0<t\leq n/2$, $l_2\in \{0,1\}^{n-t-p}$ and $l\in \{0,1\}^{n-t}$. Let $D$ be the number of classical queries and $T$ be the time of offline computations. This attack achieves a tradeoff of $T^2D=2^{2n-t}$, which is balanced at $T=D=\tilde O(2^{(2n-t)/3})$.
The remaining keys can be recovered as follows. We can recover the period $s^{l_1}=l_{13}^{-1}l_{14}k_1^{l_1}$ by solving the equation system $s^{l_1}\cdot v_i^{l_1}=0$, requiring $O(n^3)$ offline computation time. Subsequently, the remaining $t$ bits can be recovered by applying the offline Simon's algorithm on $f'(i^r,x)=f(k_2^l\|i^r, x)$, which has the period $s=k_1$ when $i^r=k_2^r$.

In the Q2 model, the construction of $F^{\mathcal L}$ is the same as in Shi {\it et al.}'s attack; see Eq.~\eqref{eq:TPP-PRFsF}. Therefore, there is no need to further partition the $(n-t)$-bit subkey $k_1^l$ into a $p$-bit part $k_1^{l_1}$ and an $(n-t-p)$-bit part $k_1^{l_2}$. The function $F^{\mathcal L}(i^l, x^{l})=\bigoplus_{u\in \mathcal L}f(i^l\| 0^{t}, x^{l} \| 0^{t} \oplus u)=G_1^{\mathcal L}(x^{l})\oplus P^{\mathcal L}(i^l, x^{l})$ has an $(n-t)$-bit period $s^{l}=l_{13}^{-1}l_{14}k_1^{l}$ when $i^l=l_{23}^{-1}l_{24}k_2^l$ or $i^l=l_{23}^{-1}l_{24}k_2^l\oplus l_{13}^{-1}l_{14}k_1^l$.
The state $|\psi_{G_1^{\mathcal L}}\rangle$ is first prepared by requiring $O\big(2^t\cdot (n-t)\big)$ quantum queries to the encryption oracle by Algorithm~\ref{al:p-xor-G2}. According to Theorem~1, we successfully recover \( l_{23}^{-1}l_{24}k^{l}_{2}\) or $l_{23}^{-1}l_{24}k_2^l\oplus l_{13}^{-1}l_{14}k_1^l$ with \( O(2^{(n-t)/2}\cdot n^3) \) offline computation time by Algorithm~\ref{al:p-xor}. The remaining key bits can be recovered using the offline Simon techniques with negligible additional cost.
\begin{figure}[H]
\centering
\includegraphics[width=15cm]{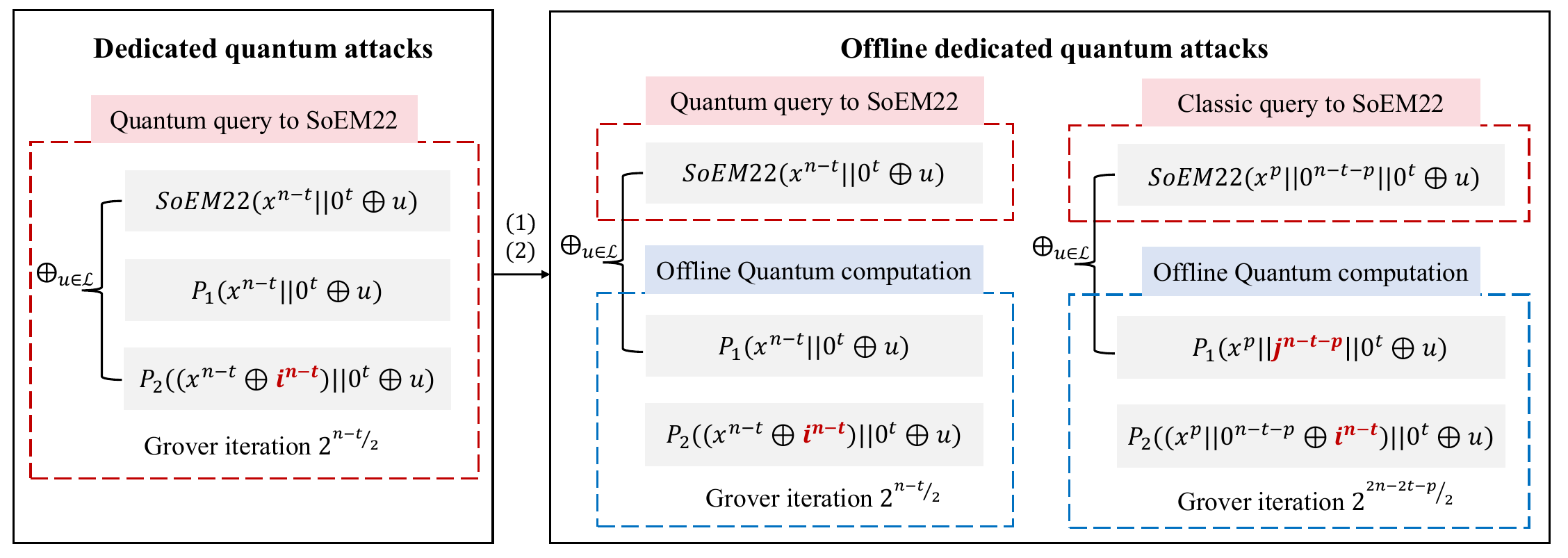}
\caption{
The overall framework of Shi {\it et al.}'s dedicated attacks~\cite{shi2024dedicated} and our offline dedicated attack on $\mathsf{SoEM22}$.}
\label{offline}
\end{figure}

To further illustrate the offline dedicated attack on TPP-PRF-based constructions, we use $\mathsf{SoEM22}$ as an example. Figure~\ref{offline} compares Shi {\it et al.}'s dedicated attack with our offline dedicated attacks in the Q2 and Q1 models. In Shi {\it et al.}'s attack, each Grover iteration requires superposition queries to the $\mathsf{SoEM22}$ oracle. In contrast, in our offline attacks, the oracle-dependent part is first extracted into the state $|\psi_{G_1^{\mathcal L}}\rangle$, and the  Grover iterations involve only offline computation based on the public permutations.

\begin{figure}[H]
    \centering
    \begin{minipage}[b]{0.45\textwidth}
        \centering
        \includegraphics[width=\linewidth]{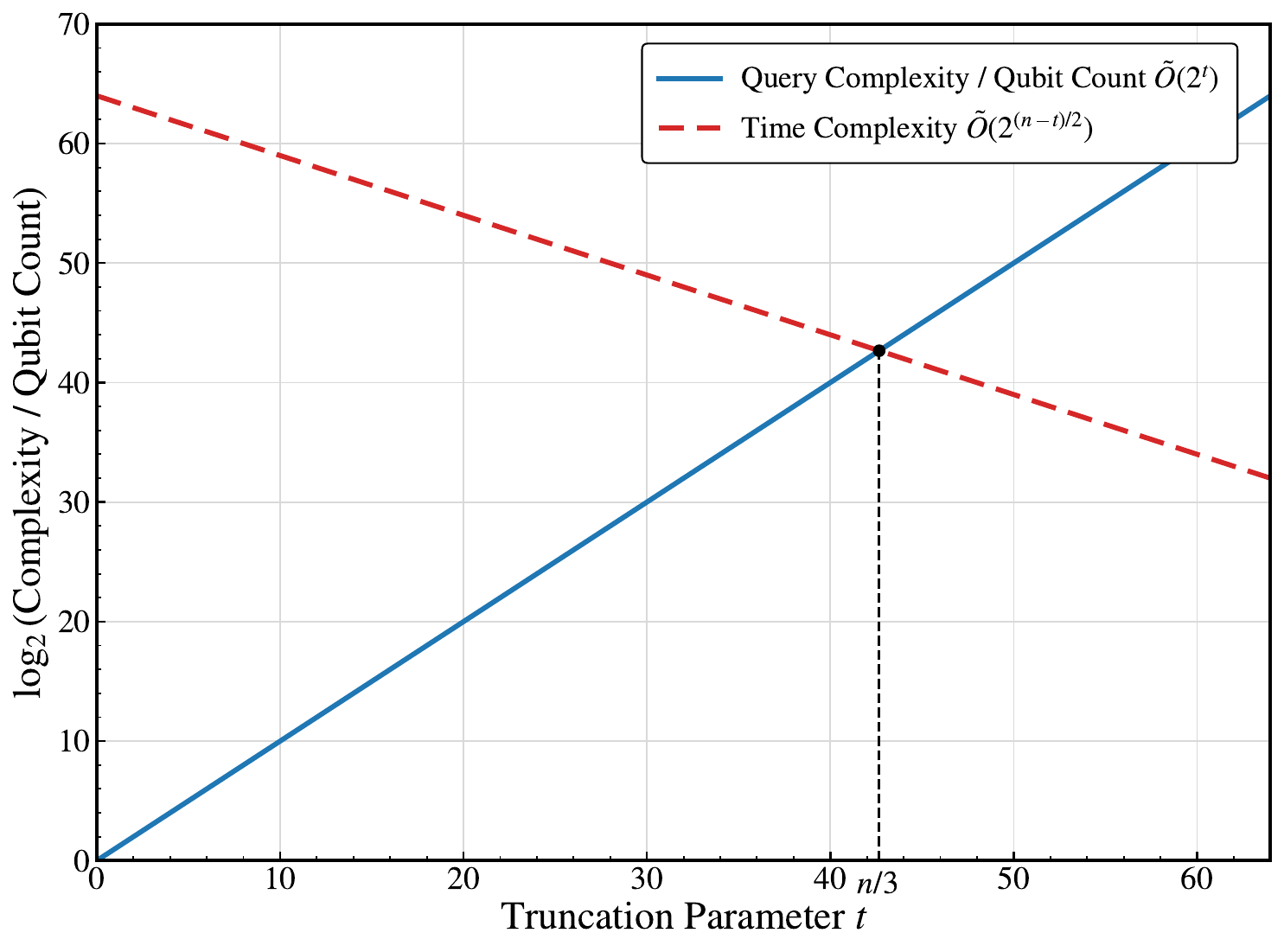}
        \vspace{0.1cm} 
        \centerline{\footnotesize(a) In the Q2 model} 
    \end{minipage}
    \hspace{0.02\textwidth}
    \begin{minipage}[b]{0.45\textwidth}
        \centering
        \includegraphics[width=\linewidth]{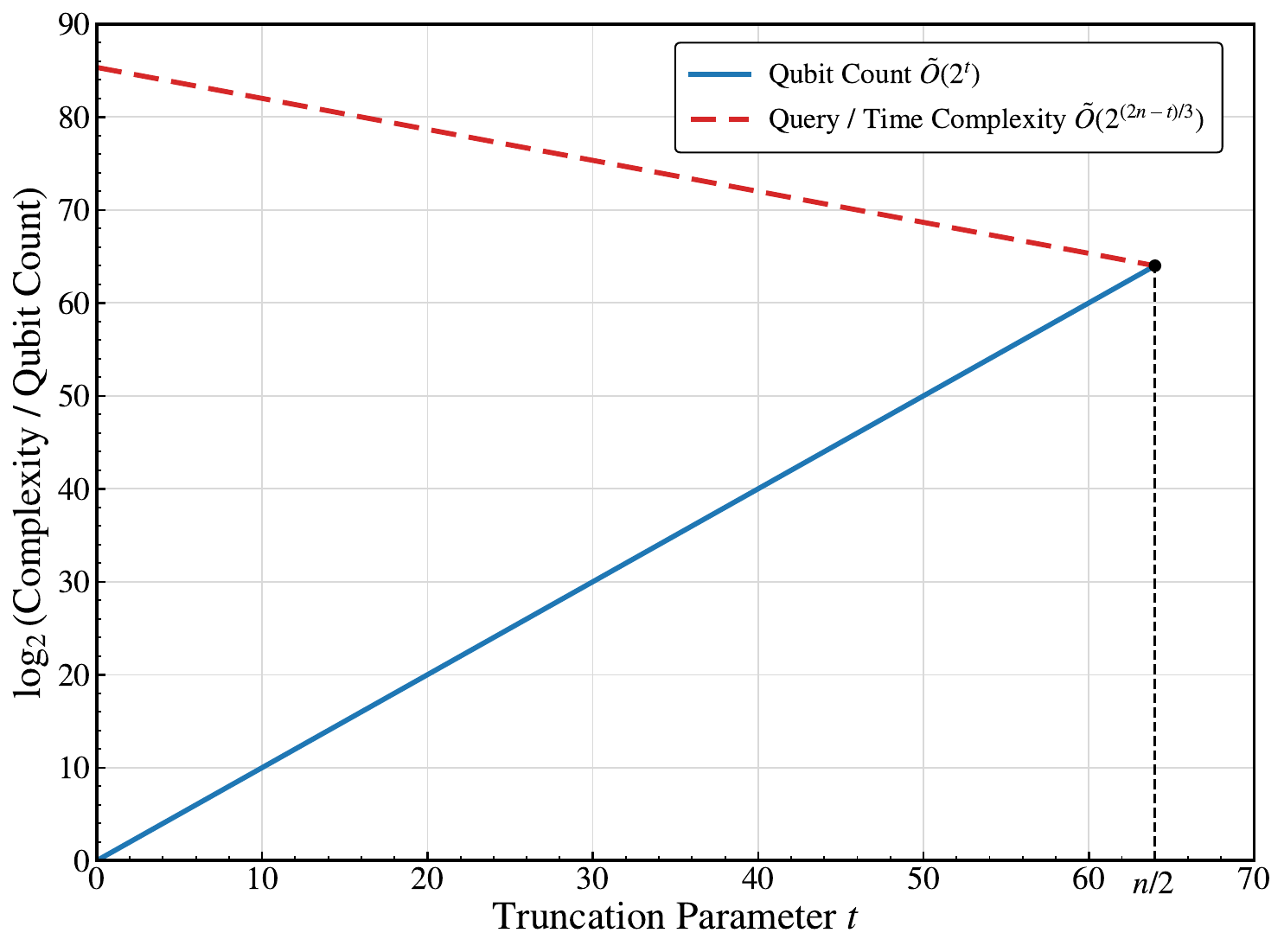}
        \vspace{0.1cm}
        \centerline{\footnotesize(b) In the Q1 model}
    \end{minipage}
    \caption{Impact of the truncation parameter $t$ on the complexities and the number of qubits for our offline attack, assuming a fixed block size of $n = 128$. The left panel (a) illustrates the resource trade-offs in the Q2 model, while the right panel (b) shows the results in the Q1 model.}
    \label{q1q2}
\end{figure}
Our optimization relies on parallelization to construct $F^{\mathcal{L}}$, which increases the required qubits by a factor of $2^t$ compared with the offline Simon's algorithm. Figure~\ref{q1q2} further shows the impact of the truncation parameter $t$ on the complexities and qubit counts for our offline attack. In the Q2 model, our attack reduces the query complexity for the same time complexity and qubit count. In the Q1 model, it reduces both query and time complexities at the cost of additional qubits.
\end{appendix}


\begin{thebibliography}{99}
\bibitem{simon1997power} Simon D. R. On the power of quantum computation. In: Proceedings 35th Annual Symposium on Foundations of Computer Science, Santa Fe, NM, USA, 1994. 116--123

\bibitem{leander2017grover} Leander G., May A. Grover meets Simon--quantumly attacking the FX-construction. In: International Conference on the Theory and Application of Cryptology and Information Security, Hong Kong, China, 2017. 161--178

\bibitem{bonnetain2019quantum} Bonnetain X., Hosoyamada A., Naya-Plasencia M., et al. Quantum attacks without superposition queries: the offline Simon's algorithm. In: International Conference on the Theory and Application of Cryptology and Information Security, Kobe, Japan, 2019. 552--583

\bibitem{shi2024dedicated} Shi T., Wu W., Hu B., et al. Dedicated quantum attacks on XOR-type function with applications to beyond-birthday-bound MACs. IEEE Trans. Inf. Forensics Secur., 2024, 19: 5971--5984

\bibitem{sun2025quantum} Sun H. W., Gao F., Xu R. X., et al. Quantum key-recovery attacks on permutation-based pseudorandom functions. IEEE Internet Things J., 2025, 12: 35692--35704

\bibitem{kim2020tight} Kim S., Lee B., Lee J. Tight security bounds for double-block hash-then-sum MACs. In: Annual International Conference on the Theory and Applications of Cryptographic Techniques, Zagreb, Croatia, 2020. 435--465

\bibitem{chen2019build} Chen Y. L., Lambooij E., Mennink B. How to build pseudorandom functions from public random permutations. In: Annual International Cryptology Conference, Santa Barbara, CA, USA, 2019. 266--293

\bibitem{bellare1998luby} Bellare M., Krovetz T., Rogaway P. Luby-Rackoff backwards: Increasing security by making block ciphers non-invertible. In: International Conference on the Theory and Applications of Cryptographic Techniques, Espoo, Finland, 1998. 266--280

\bibitem{guo2019beyond} Guo C., Shen Y., Wang L., et al. Beyond-birthday secure domain-preserving PRFs from a single permutation. Des. Codes Cryptogr., 2019, 87: 1297--1322
\bibitem{bhattacharjee2022cencpp} Bhattacharjee A., Dutta A., List E., et al. Cencpp*: beyond-birthday-secure encryption from public permutations. Des. Codes Cryptogr., 2022, 90: 1381--1425

\end{thebibliography}

\end{document}